# Efficient photocatalytic nitrogen fixation from air under sunlight via iron-doped WO$_3$[1]

Yuanfang Shen [a], Jingxuan Shou [b], Liangchen Chen [a], Weihang Han [a], Luping Zhang [a], Yutong Chen [a], Xuewei Tu [a], Shangfu Zhang [a], Qiang Sun [c,†], Yurong Chang [d], Hui Zheng [a,*]

[a] College of Material, Chemistry and Chemical Engineering, Hangzhou Normal University, Hangzhou 311121, P.R. China

[b] College of civil Engineering, Zhejiang University of Techonlogy, Hangzhou 311121, P.R. China

[c] Australian Research Council Centre of Excellence for Nanoscale BioPhotonics (CNBP), School of Science, RMIT University, Melbourne, VIC 3001, Australia

[d] Henan Qite New Material Co., Ltd., Zhengzhou 450000, P.R. China

[†] Corresponding author: qiang.sun@rmit.edu.au

[*] Corresponding author: huizheng@hznu.edu.cn

## Abstract:

Photocatalytic nitrogen fixation from air directly under sunlight can contribute significantly to carbon neutralization. It is an ideal pathway to replace the industrial Haber Bosch process in future. A Fe-doped layered WO$_3$ photocatalyst containing oxygen vacancies was developed which can fix nitrogen from air directly under sunlight at atmospheric pressure. The iron doping enhances the transport efficiency of photogenerated electrons. The photocatalytic efficiency is around 4 times higher than that of pure WO$_3$. The optium nitrogen fixation conditions were examined by orthogonal experiments and its nitrogen fixation performance could reach up to 477 $\mu g \cdot g_{cat}^{-1} \cdot h^{-1}$ under sunlight. In addition, the process of nitrogen fixation was detected by situ infrared, which confirmed the reliability of nitrogen fixation. Also, modelling on the interactions between light and the photocatalyst was carried out to study the distribution of surface charge and validate the light absorption of the photocatalyst. This work provides a simple and cheap strategy for photocatalytic nitrogen fixation from air under mild conditions.

## 1. Introduction:

78% of air is nitrogen, which seems that there are inexhaustible nitrogens ready for us to use. Nevertheless, due to the high bond energy of nitrogen (940.95 kJ/mol), it is not easy to activate and it has always been considered as an inert gas[1]. The conversion of dinitrogen to ammonia is a very important carbon-neutral reaction[2, 3]. Traditionally, the Haber Bosch method was employed for nitrogen fixation with high temperature and high-pressure in industry since last century[4]. This thermal nitrogen fixation requires burning a large number of fossil fuel, which will cause irreversible damage to environment[5]. The energy consumption of synthetic ammonia in industry is as high as 2.5 EJ/yr and the carbon emission is as high as 340 Mt CO$_2$ eq/yr, which is the most significant energy consumption among chemical commodities[6]. Therefore, developing a green method for nitrogen fixation is in imperious demands. Sunlight-driven nitrogen fixation under normal temperature and pressure has become one of the ideal pathways. Due to the high chemical energy barrier of nitrogen, photocatalytic nitrogen fixation always has its unique bottleneck[7]. Firstly, the complete conversion from nitrogen to ammonium needs 6 electrons which needs quite a few free electrons. However, light-excited electron-hole pairs are very easy to recombine, and electron lifetime is not long enough (hundreds ns)[8]. It requires photocatalysts to have good ability to generate and transport electrons. Secondly, the surface of photocatalyst needs to have sufficient active sites to adsorb nitrogen. Generally, active sites like N, O, S and other vacancies can effectively capture photoelectrons[9-11]. Many catalysts have weak nitrogen adsorption, which will directly affect the nitrogen fixation performance of the catalyst.

Indian soil scientists discovered the first case of photocatalytic nitrogen fixation in the desert, and later re-presented by Schrauzer and Guth[12],[13]. They found that transition metal iron-doped $TiO_2$ as a catalyst does have nitrogen fixation properties[14]. Due to the wide band gap of $TiO_2$, its light absorption efficiency is low ($Eg \approx 3.2eV$)[15]. As such, these semiconductor substrates photocatalyst with a narrow band gap beccome a research hotspots to enhance the visible light harvesting efficiency such as $WO_{3-x}$[16-18], BiOX[19, 20], g-$C_3N_4$[21], etc. It is worth mentioning that $WO_{3-x}$ has local surface plasmon resonance, which makes it has absorbtion in the near-infrared region[22]. Once the plasmon resonance is excited, it will concentrate and locally amplify the light absorption on the catalyst surface, making $WO_{3-x}$ exhibit fascinating light conversion efficiency[23]. In addition, vacancy engineering is also a common method for designing photocatalysts[7, 11]. Due to its unique crystal structure, $WO_3$ can maintain a certain amount of oxygen vacancies. A new energy- deficient band can be generated below the conduction band when the oxygen vacancy increases to a certain amount[24]. This energy band makes the catalyst produce a new absorption peak between 400-700nm. Compared with bulk $WO_3$, 2D flake appearance $WO_3$ significantly can increase the interface contact area and charge transfer efficiency[17, 25, 26]. This makes the flake $WO_3$ has more photocatalytic active.

Herein, in this work, the rich and cheap transition metal Fe is designed to incorporate into the matrix $WO_3$ to form a Fe-$WO_3$ photocatalyst with a flake appearance. There are a large number of electron-rich oxygen vacancies on the surface of Fe-$WO_3$ and nitrogen in air could be adsorbed on these vacancies which can realize photocatalytic nitrogen fixation. Electrons are injected from the vacancies into the anti-bonding orbital of N≡N and activate nitrogen[27]. The doping of Fe can improve electron transmission and light absorption efficiency, thereby improving the catalytic performance. The temperature, catalyst dosage and other conditions of fixation nitrogen were optimized by orthogonal experiments. The in-situ infrared experiment was performed to verify occurrence of nitrogen fixation reaction from air directly. It has been found that the nitrogen fixation efficiency of Fe-$WO_3$ can reach 477 $\mu g \cdot g_{cat}^{-1} \cdot h^{-1}$, which is 4 times higher than that of pure $WO_3$. This work provides a new pathway for photocatalytic nitrogen fixation under ambient temperature and pressure.

## 2. Experimental section

## 2.1 Materials and reagents

The chemical reagents are all of the analytical grades and can be used directly without further purification. $WS_2$ (99.9%) and $FeCl_3$ (98%) were purchased from Aladdin. Sodium citrate (>99%) and sodium hydroxide (>96%) were purchased from General-Reagent Company. Sodium nitroferricyanide dehydrate (99%) was purchased from Jiuding Chemical Technology Co., Ltd. Sodium hypochlorite was purchased from Sinopharm Chemical Reagent Co., Ltd. Salicylic acid (99.5%) was purchased from Macleans Reagent Co., Ltd. The 0.22μm filter was purchased from Titan Experimental Company.

## 2.2 Preparations of the catalysts

1 mmol of tungsten disulfide was dispersed in 30 mL of pure water, followed by ultrasonic peeling for 4 hours. 0.25 mmol $FeCl_3$ was dissolved in 10 mL of pure water which was then added dropwise to the ultrasonic tungsten disulfide suspension. In a 100 °C water bath, the pure water was evaporated to dry while stirring. In the final step, the precursor was placed in a muffle furnace and calcined at 500°C for 3 hours to obtain an iron-doped tungsten oxide catalyst. The catalyst is named X Fe-$WO_3$ in which X is the molar amount of $FeCl_3$. After optimization, X=0.25 is the best ratio for this nitrogen fixation.optium experimenta.

## 2.3 Determination of ammonium

The salicylic acid was prepared into a 50g/L alkaline solution with sodium hydroxide and sodium citrate. 20 μL of 10 g/L sodium nitroferricyanide solution and 20 μL of c (NaClO) 0.05 mol/L sodium hypochlorite solution were prepared. The reaction solution was filtered through a 0.22 μm filter to obtain the reaction solution. The 0.5mL reaction solution was diluted to 2 mL which was added to the prepared indicators: 100 μL of the salicylic acid solution, 20 μL of sodium nitroferricyanide solution, and 20 μL of sodium hypochlorite solution. After shaking, we let it stand for 1h. Then, we measured the absorbance at a wavelength of 697.5 nm with an ultraviolet-visible spectrophotometer as demonstrated in literature[28]. The ammonium concentration was calculated according to the standard curve.

## 2.4 Characterizations

X-ray diffraction (XRD, D8 Advance) with Cu Kα radiator was used to determine the crystal configuration of catalysts. Scanning electron microscope (SEM-EDS, Zeiss Sigma 500) and high-resolution transmission electron microscopy (HRTEM, FEI Tecnai G2 F20) were used to detect the surface morphology and structure analysis of catalysts. X-ray photoelectron spectroscopy (XPS, Thermo Scientific K-Alpha) with Al Kα rays was used to detect the chemical state of the catalyst surface. The free radicals and vacancies generated by the catalysts were measured by electron spin resonance (ESR, Bruker A300). The impedance of the catalyst was tested at a frequency of 0.01-100kHz in a 2.5mmol/L potassium ferricyanide solution, and tested under dark and light (EIS, CHI750). For the electrochemical measurement, the system adopted a three-electrode configuration. Platinum wire was used as the counter electrode, and saturated calomel was used as the reference electrode. The catalyst was deposited on the ITO as a working electrode. The photocurrent response test was performed in a 0.5 mmol/L sodium sulfate solution. The light absorption of the catalyst was measured with an ultraviolet-visible diffuse reflectance spectrophotometer (DRS, UV-3600). The photoluminescence after the reaction was analyzed by a fluorescence chromatograph (PL, F-7000). Radical lifetimes were measured with steady-state/transient fluorescence spectrometers (FLS 980). In situ infrared (in-situ FTIR, Bruker TENSOR II) detection of the product produced by the catalyst in the aqueous solution was carried out for real-time detection, thereby inferring the process of the nitrogen fixation reaction. The laser power meter (VLP-2000-10W) was used to measure the intensity of sunlight and ultraviolet light (UV).

## 2.5 Nitrogen photo fixation reaction

In a typical experiment, 10 mg of catalyst was dispersed in 10 mL of pure water. The entire reaction system was bubbled with air and stirred at 700 rpm for 30 minutes in dark. 0.6 mL initial liquid was taken as the sample to reference. Then the reaction system was put under the light source, and 0.6 mL of the reaction solution was taken out every hour to examine. The ammonium concentration was detected by the indophenol blue method.

The catalyst operation steps after 5th cycles are as follows: 10 mg of the freshly prepared catalyst is dispersed in 10 mL water, bubbled with air and illuminated for 5 hours. The catalyst was collected by centrifugation and washed three times with pure water. The obtained catalyst was counted as 1th cycle. After repeating this operation 4 times, the catalyst was detected.

## 3. Results and discussions

## 3.1 Structure and morphology research

A series of catalysts were prepared by above method. X-ray diffraction (XRD) was used to determine the phase of catalysts (Figure 1). After the ultrasonic blending of iron and $WS_2$, the crystal phase does not change, which is highly consistent with the raw $WS_2$. It means that iron did not enter the $WS_2$ lattice before calcination. Nevertheless, the crystal structure of catalysts has changed significantly after calcination. By comparing with the PDF standard card, it is determined that the crystalline form of the calcined catalyst is consistent with 75-2072, which is $WO_3$. This indicates that the tungsten element is oxidized by air after calcination to form $WO_3$. The doping of Fe leads to the reduction of diffraction peaks of certain crystal planes, such as (0 2 0), (2 0 0), (0 2 1), (2 2 0). This means that Fe was successfully doped into the $WO_3$ crystal lattice, which affects the diffraction of the $WO_3$ crystal form. It is also intuitively recognized from Zone I (Figure 2e).

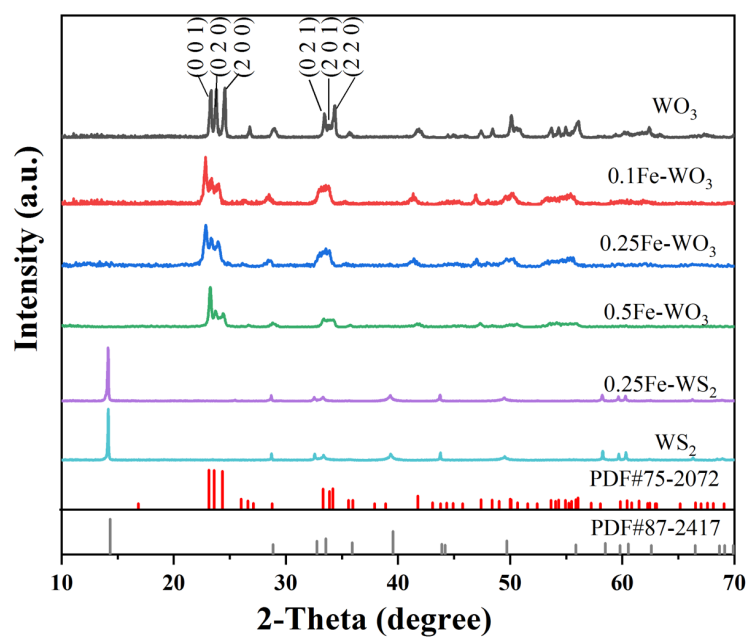

Figure 1. XRD patterns of tungsten sulfide doped with different proportions of iron.

From the morphology of the catalyst surface observed by scanning electron microscope, the surface of 0.25Fe-WS$_2$ is smooth and morphology is a hexagonal disk after ultrasonic stripping (Figure 2a, b). After calcination, the overall morphology of the entire catalyst is uniform as a disk and still in the hexagonal shape, while its surface becomes rough. The high magnification image showed that many small bumps about 30 nm in diameter appear on the catalyst's surface (Figure 2c, d). The sulfur in WS$_2$ is volatilized after being oxidized by air, and tungsten is oxidized to WO$_3$. Therefore, the catalyst keeps the hexagonal morphology, but the phase changes completely. To further determine the structure information of 0.25Fe-WO$_3$, the HRTEM test was carried out (Figure 2e, f). A variety of apparent lattice fringe spacings were found through the analysis of lattice fringes. And these lattice spacings match precisely with the XRD diffraction information, which proves the successful synthesis of the catalyst. Distortions such as lattice dislocations appeared in many places in the Zone I of the HRTEM image, which is a good proof of the successful doping of Fe. Fe atomic radius is smaller than tungsten, resulting in local lattice mismatch. Eventually, there were many short-range disorder phenomena. At the same time, the existence of lattice defects will generate a large number of oxygen vacancies. This vacancy will provide sufficient active sites for nitrogen adsorption.

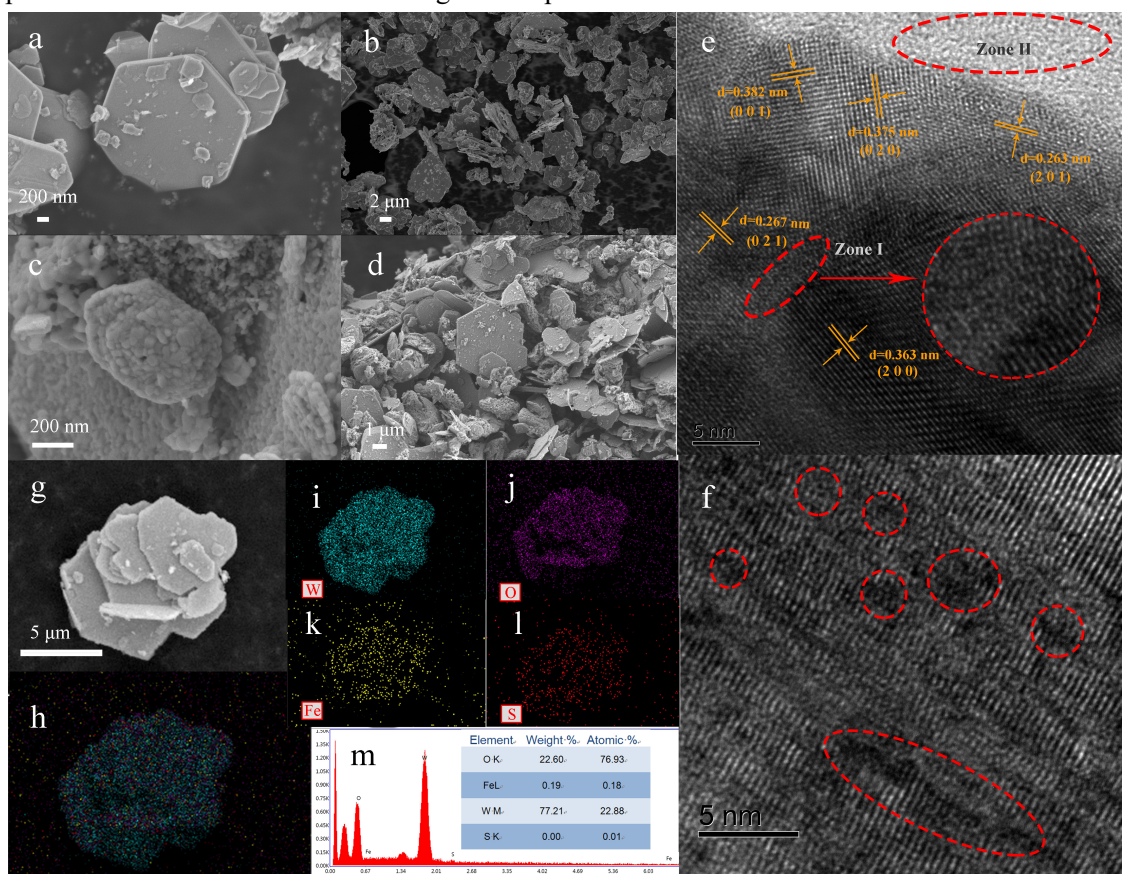

Figure 2. SEM of 0.25Fe-WS$_2$(a, b), 0.25Fe-WO$_3$ (c, d), HRTEM (e,f), mapping(g-l) and EDS (m) of 0.25Fe-WO$_3$.

Part of the iron cannot be successfully incorporated into WO$_3$ due to lattice mismatch, a disordered structure will appear on the surface of the catalyst (zone II)[16]. Figure 2(g-l) shows the distribution of elements on the surface of 0.25Fe-WO$_3$. Fe, O, W, S elements were uniformly distributed in the catalyst, proving the successful preparation of 0.25Fe-WO$_3$. EDS detected the element content of 0.25Fe-WO$_3$, and it was found that the surface was mainly W, O and a small amount of Fe (Figure 2m). It indicates that iron is uniformly doped into the catalyst surface and inside. That is consistent with the absence of characteristic peaks for iron in the XRD pattern.

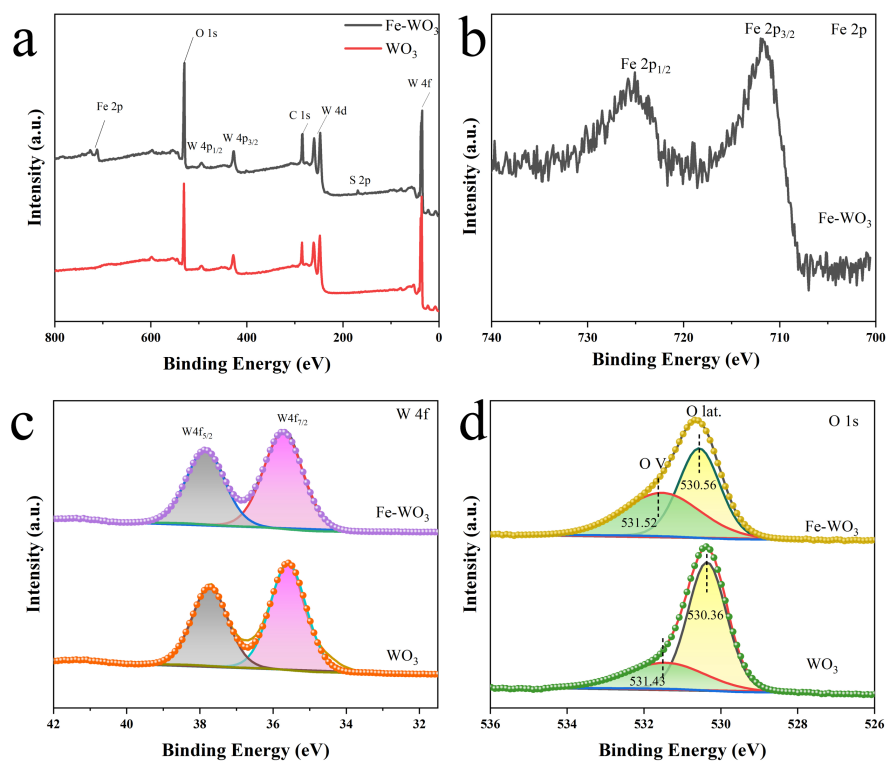

Figure 3. XPS image of 0.25Fe-WO$_3$ (a) full spectrum, (b) Fe 2p, (c) W 4f and (d) O1s.

To determine the chemical composition of 0.25Fe-WO$_3$ and WO$_3$, and compare the chemical state, catalysts were analyzed by X-ray photoelectron spectroscopy (XPS). All XPS spectra were firstly calibrated with C 1s peaking at 284.8eV. The full spectrums determine the catalysts' element composition (Figure 3a). Fe-WO$_3$ contains element Fe, W, O, and S. Figure 3b shows that the binding energies of Fe 2p$_{1/2}$ and Fe 2p$_{3/2}$ are at 725.36 eV and 711.76 eV, respectively, which proves the presence of Fe (III)[29]. It demonstrates that Fe was successfully incorporated into the WO$_3$ lattice. In the spectral peaks of XPS W 4f (Figure 3c), the binding energies at 37.9 eV and 38.5 eV correspond to the polar nuclear peaks W4f 5/2 and W4f 7/2, respectively[30]. It confirms the existence of WO$_3$. There are a lot of lattice oxygen (O lat.) and oxygen vacancies (O V) in the O 1s spectrum (Figure 3d)[29]. With the incorporation of Fe, the content of oxygen vacancies obviously increases. This indicates that the incorporation of Fe will aggravate the distortion of the catalyst lattice and increase the oxygen vacancy content. After nitrogen enters the aqueous solution, it is easily absorbed by the electron-rich oxygen vacancies on the catalyst. The electrons on the oxygen vacancies are quickly injected into the anti-bonding orbital of the nitrogen triple bond to activate the nitrogen[27].

## 3.2 Optical properties

In order to determine the electron transfer efficiency in catalyst, the impedance of catalysts

Fe-WO$_3$ and WO$_3$ under light and dark conditions was tested by the Nyquist curve (Figure 4a). By comparing the radius of semicircle in high-frequency regions, it is concluded that the impedance of Fe-WO$_3$ is smaller than that of WO$_3$ under the same conditions[31, 32]. The metallic properties of Fe facilitate the generation of electrons when excited by light (Figure 4b). Therefore, the incorporation of Fe can significantly reduce the charge transfer resistance and enhance the charge separation efficiency. Also, the impedance of the same catalyst under light conditions is lower than dark, indicating that light can further improve the electron transport efficiency[33].

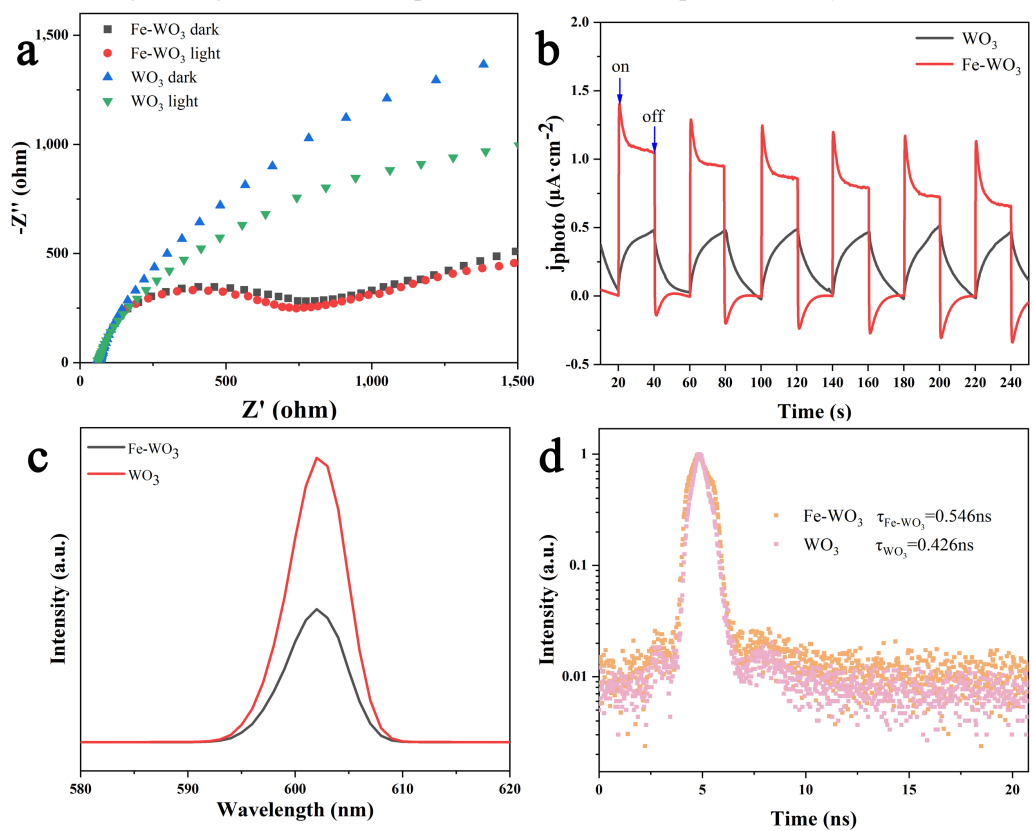

Figure 4. Nyquist curve (a), Transient current (b), steady-state PL spectra (c) and transient fluorescence lifetime decay curves (d) of Fe-WO$_3$ and WO$_3$.

To further explore the optical properties of catalysts, bandgap position and forbidden bandwidth were tested. Figure 5a shows the absorption spectra of catalysts WO$_3$ and 0.25 Fe-WO$_3$ under ultraviolet-visible light. The incorporation of iron greatly improves the absorption efficiency of 0.25 Fe-WO$_3$ in the visible light region. 0.25 Fe-WO$_3$ can also generate electron-hole pairs under longer wavelength excitation, which improves the effectiveness of light utilization.

As shown in Figure 4e, both catalysts WO$_3$ and Fe-WO$_3$ show strong emission peaks around 603 nm. It indicates that the photo-induced charge carriers recombine a lot at this wavelength. After iron was incorporated into WO$_3$, the luminescence intensity of the catalyst decreased significantly. It is proved that Fe can effectively inhibit the recombination of photogenerated electron-hole pairs. In addition, it was also determined kinetically that Fe incorporation enhanced the electron existence time by measuring transient fluorescence lifetime decay curves (Figure 4d).

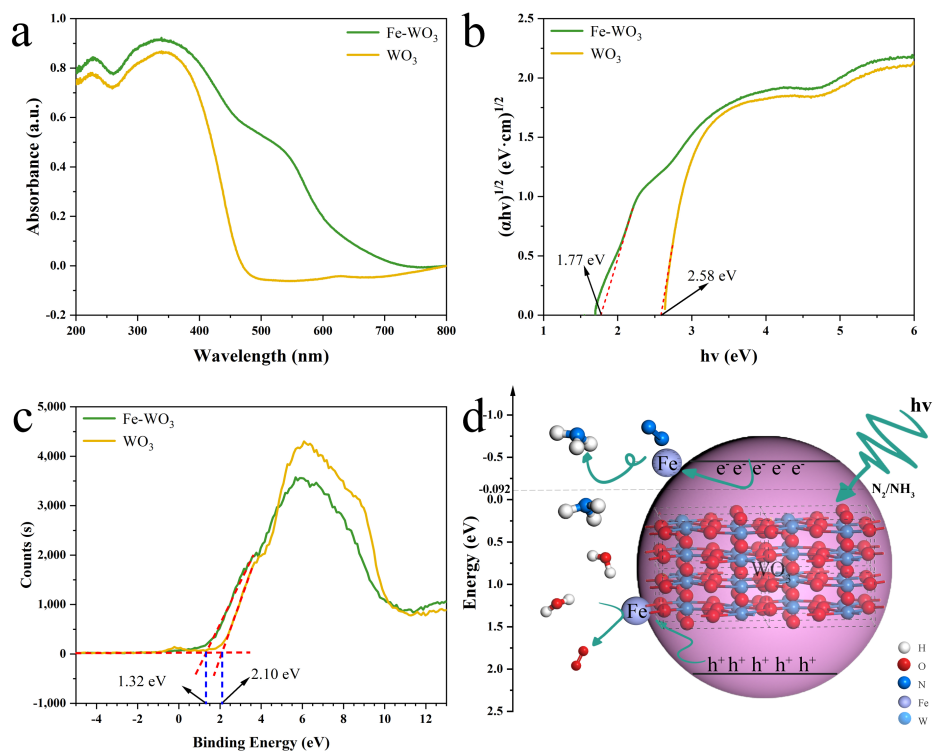

Figure 5. Light absorption of 0.25 Fe-WO$_3$ (a), DRS (b), Eg value (c) and XPS-VB (d).

The bandgap width (Eg) of catalyst is calculated from the formula[34]:
$$\alpha h\upsilon = C(h\upsilon - E_g)^2$$

The bandgap distance directly affects the efficiency of photocatalysis of the catalyst to generate electron-hole pairs. The narrower the band gap, the easier it is to excite electron-hole pairs. Figure 5b shows that the band gap of 0.25 Fe-WO$_3$ is lower than that of WO$_3$ which it is easier to generate electron-hole pairs under light. The valence band (VB) of catalysts was determined by XPS-VB (Figure 5c). According to the forbidden band width as the gap between VB and conduction band (CB), the value of CB can be obtained[35]. The position of CB directly determines the reduction ability of the catalyst, which is closely related to the possibility of nitrogen fixation. The reduction potential of the catalyst needs to be lower than the potential required for nitrogen fixation (-0.092eV)[36], so that the nitrogen fixation process can happen smoothly. The reduction potentials of the catalysts Fe-WO$_3$ and WO$_3$ are -0.45eV and -0.48eV, respectively. Therefore, from a thermodynamic point of view, both catalysts can achieve nitrogen fixation. The addition of Fe greatly improves the light absorption efficiency of Fe-WO$_3$ for its narrow band gap (Figure 5d). At the same time, the underfilled 3d orbital of Fe is conducive to electron transport[37], so that Fe-WO$_3$ has a higher nitrogen fixation efficiency (Figure 8b).

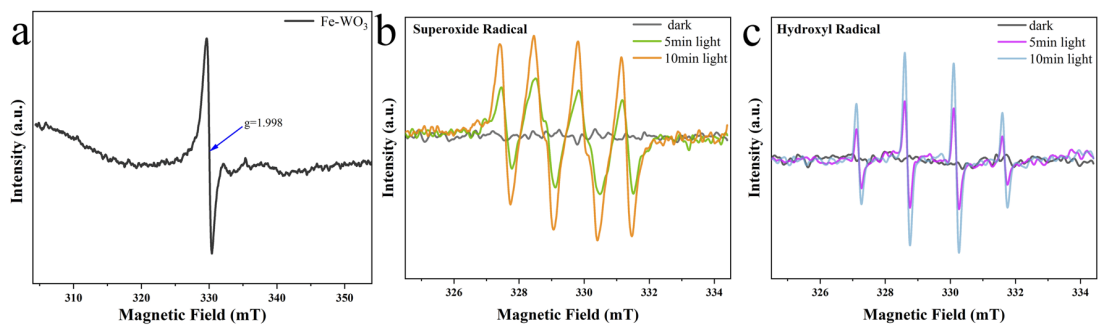

Figure 6. ESR spectrum of 0.25Fe-WO$_3$ g-value (a), superoxide radical (b) and hydroxyl radical (c).

Electron spin resonance spectrometer (ESR) was used to detect unpaired electrons in samples. It was detected that the g-value of 0.25Fe-WO$_3$ shows a strong signal at 1.998[38], which indicates that the catalyst surface contains a large number of vacancies. Compared with the O 1s of XPS data, it is presumed to be oxygen vacancies (Figure 6a). Oxygen vacancies serve as active sites where nitrogen can be easily adsorbed[27]. Then nitrogen activation can be achieved by electron input to the nitrogen antibonding orbital. Therefore, the content of oxygen vacancies will directly affect the efficiency of nitrogen fixation. The capture of free radicals by DMPO (dimethyl pyridine N-oxide) confirms the existence of superoxide free radicals and hydroxyl free radicals in methanol and aqueous solution, respectively. Both of them cannot be detected in dark. When light is turned on, superoxide radicals and hydroxyl radicals are produced at the same time. And with the prolongation of light, the concentrations of both free radicals increase. The results directly confirm the light-driven generation of free radicals (Figure 6b,c). The generation of hydroxyl radicals indicates that the catalyst can consume holes (h$^+$) under illumination. It can be predicted that the catalyst can achieve nitrogen fixation without introducing additional sacrificial agents[39]. The conjecture was confirmed in orthogonal experiments. Oxygen combines with electrons to produce superoxide radicals. The reduction potential required for the nitrogen fixation reaction to occur is lower than that generated by superoxide radicals (-0.33 eV). Therefore, it can be confirmed that the catalyst can achieve nitrogen fixation thermodynamically[40].

The whole process is hypothesized in Figure 7. The catalyst is photoexcited to generate electron-hole pairs in the first step. Electrons are transported through the transition metal to holes on the catalyst surface. Then electrons are injected into the anti-bonding orbital of nitrogen to achieve nitrogen activation. The proton hydrogen in the aqueous solution is gradually connected to the nitrogen, and the bond is broken to form ammonium radical. On the other hand, The catalyst oxidation potential is not sufficient to directly oxidize water to form hydroxyl radicals. Actually, a large number of hydroxyl radicals were detected in the aqueous solution. It may be due to the reaction of Fe$^{3+}$ with holes. In the final step, the hydroxyl groups further occupy the holes to generate water and oxygen[18, 41-43].

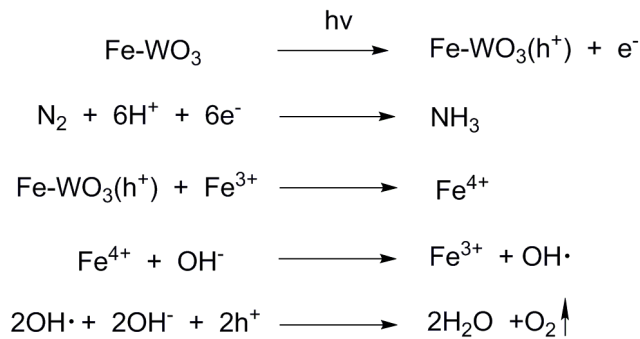

$$Fe\text{-}WO_3 \xrightarrow{h\nu} Fe\text{-}WO_3(h^+) + e^-$$

$$N_2 + 6H^+ + 6e^- \longrightarrow NH_3$$

$$Fe\text{-}WO_3(h^+) + Fe^{3+} \longrightarrow Fe^{4+}$$

$$Fe^{4+} + OH^- \longrightarrow Fe^{3+} + OH\cdot$$

$$2OH\cdot + 2OH^- + 2h^+ \longrightarrow 2H_2O + O_2\uparrow$$

Figure 7. The possible nitrogen fixation pathway.

## 3.3 Nitrogen fixation performance

To determine the nitrogen fixation efficiency of this photocatalyst, a series of photocatalytic fixation nitrogen experiments from air directly were carried out in water. The indophenol blue indicator method was used to detect the concentration of ammonium ions, and the standard curve is shown in Figure 8a. In Figure 8b, the nitrogen fixation performance of catalysts with different iron doping levels was compared. Interestingly, when the amount of Fe material is 0.1 times and 0.25 times of tungsten, the synergistic effect of iron is fully presented. The photocatalytic activity is about 4 times higher than the nitrogen fixation activity of pure $WO_3$. It well shows that a specific concentration of iron content can significantly improve the charge separation efficiency and electron transport capacity, which is beneficial to the utilization of light and the activation of nitrogen by catalysts. At the same time, a parallel test was carried out on $0.25Fe\text{-}WO_3$ under argon atmosphere. The experimental results show that this catalyst hardly has any nitrogen fixation activity under argon atmosphere. In addition, it was confirmed by experiments that photocatalytic activity of the catalyst $0.25Fe\text{-}WS_2$ was slightly lower than $WO_3$. This phenomenon eliminates the influence of the incomplete reaction product of the precursor on the performance test. It can be proved that $0.25Fe\text{-}WO_3$ can convert nitrogen into ammonium from air directly.

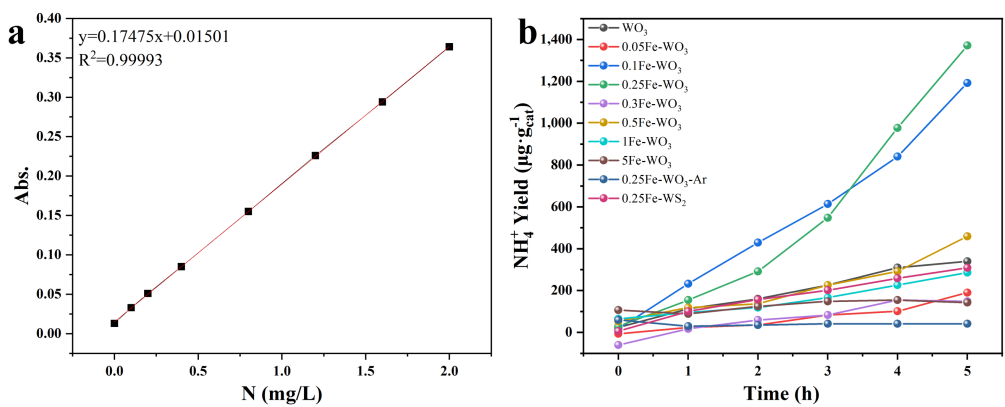

Figure 8. Indophenol blue standard curve (a) and nitrogen fixation effect (b)

The optimal combination of nitrogen fixation conditions for $0.25Fe\text{-}WO_3$ was determined through orthogonal experiments. Three conditions of nitrogen fixation temperature (A, °C), catalyst dosage (B, mg), and sacrificial agent concentration (C, mol/L) have been optimized. The optimization table is shown in Table 1. Considering that the ammonia gas dissolved in the aqueous solution will escape with the temperature rises, as such the maximum temperature is set to 55 °C.

Table 1 Optimization of nitrogen fixation factors

| A(temperature) | B(catalyst dosage) | C(Sacrificial agent concentration) |
|---|---|---|
| 15 | 2 | 0 |
| 35 | 5 | 1 |
| 55 | 8 | 5 |

Table 2 Orthogonal experiment results

| Entry | A | B | C | NH$_3$ Yield[a] ($\mu g \cdot g_{cat}^{-1} \cdot h^{-1}$) |
|---|---|---|---|---|
| 1 | 1 | 1 | 1 | 215 |
| 2 | 1 | 2 | 2 | 71 |
| 3 | 1 | 3 | 3 | 33 |
| 4 | 2 | 1 | 2 | 171 |
| 5 | 2 | 2 | 3 | 111 |
| 6 | 2 | 3 | 1 | 27 |
| 7 | 3 | 1 | 3 | 196 |
| 8 | 3 | 2 | 1 | 106 |
| 9 | 3 | 3 | 2 | 67 |

[a] Add certain amount catalyst to 10ml of aqueous solution, stir for 30min in the dark, and then measure ammonium concentration after 5h of light

Table 3 Range analysis of orthogonal experiment results

| K/$\mu g \cdot g_{cat}^{-1} \cdot h^{-1}$ | A | B | C |
|---|---|---|---|
| K$_1$ | 319 | 491 | 348 |
| K$_2$ | 309 | 288 | 308 |
| K$_3$ | 369 | 127 | 340 |
| R | 61 | 364 | 39 |

By analysizing the extreme value (R) of the orthogonal result, it is determined that the amount of catalyst has the greatest impact on the nitrogen fixation. The impact of the reaction temperature is less significant than the amount of catalyst, and the effect of the concentration of sacrificial agent is negligible. In a short summary, the orthogonal experiment table shows that the optimal condition for nitrogen fixation reaction of 0.25Fe-WO$_3$ is to add 2 mg of catalyst into 10 ml of pure water at 55℃.

After obtaining the optimal catalyst, the light sources were optimized too. The nitrogen fixation experiments were carried out with sunlight and UV light as light sources under orthogonal optimal conditions (Figure 9). The catalyst 0.25Fe-WO$_3$ can achieve nitrogen fixation under both UV and sunlight. The rate of nitrogen fixation under sunlight is much higher than that under the UV light, and the highest nitrogen fixation rate can reach 477 $\mu g \cdot g_{cat}^{-1} \cdot h^{-1}$ under sunlight. It can be concluded that 0.25Fe-WO$_3$ can realize better nitrogen fixation under sunlight. In addition, the nitrogen fixation rate is affected by the light intensity. The higher light intensity, the higher nitrogen fixation efficiency.

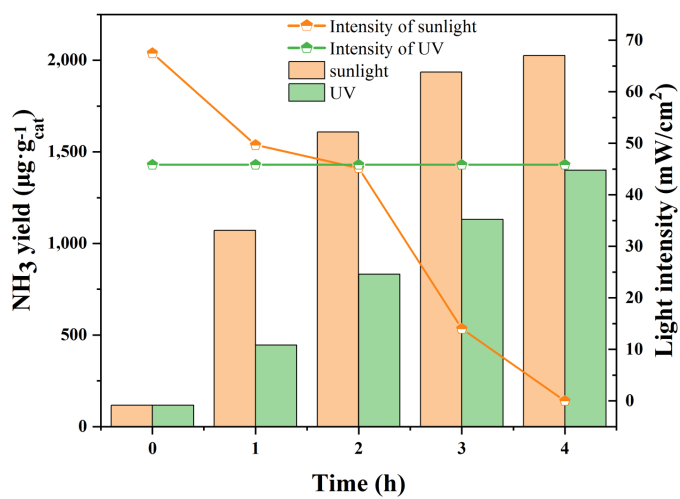

Figure 9. Comparison of nitrogen fixation effects under different light sources and the sunlight experiment was performed outside from 13:20 to 17:20 (Hangzhou, eastern longitude 120°01′, northern latitude 30°29′) on 12-Jan-2022.

In order to verify the structural stability of catalysts, the crystal structures and surface morphologies of freshly prepared catalyst and after 5th cycles were compared (Figure 10) The lattice structure of catalysts did not change after 5 cycles. It is still iron-doped $WO_3$ crystal form. And after cycles, the surface of catalyst showed many small-particle raised hexagonal flakes, which was not much different from that of freshly prepared catalyst. This proves that the catalyst has good structural stability.

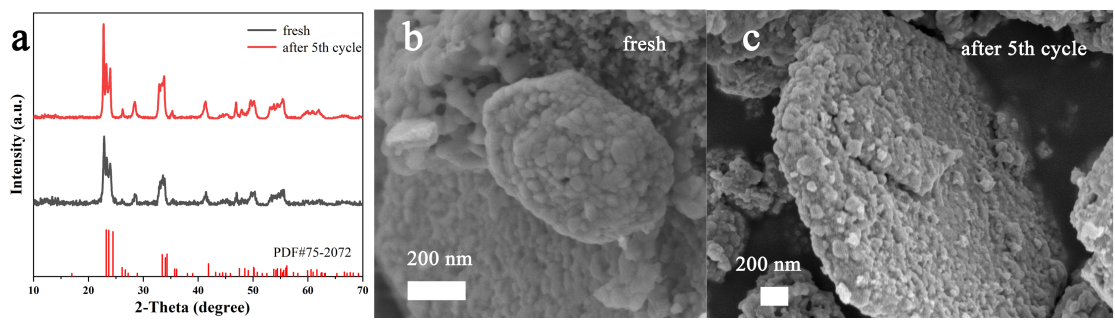

Figure 10. XRD patterns(a) and SEM images(b,c) of the freshly prepared catalyst and catalyst after 5th cycle

## 3.4 Nitrogen activation mechanism

In order to explore the nitrogen fixation process, time-dependent in-situ FTIR measurements were performed. As shown in Figure 11, the absorption peaks at wavelengths of $3441 cm^{-1}$ and $1424 cm^{-1}$ correspond to the stretching and deformation of the N-H peak, respectively. The peak at $3555 cm^{-1}$ is attributed to the O-H stretching vibration corresponding to the adsorbed water. The peak of $1629 cm^{-1}$ represents the nitrogen absorbed in the aqueous solution. With the prolongation of illumination time, the nitrogen content in the aqueous solution increases, and the $NH_4^+$ and $NH_3$ content at $2964 cm^{-1}$ and $1293 cm^{-1}$, respectively, also increase gradually. It is confirmed that nitrogen can be efficiently absorbed from air directly under this system.[2, 44-46].

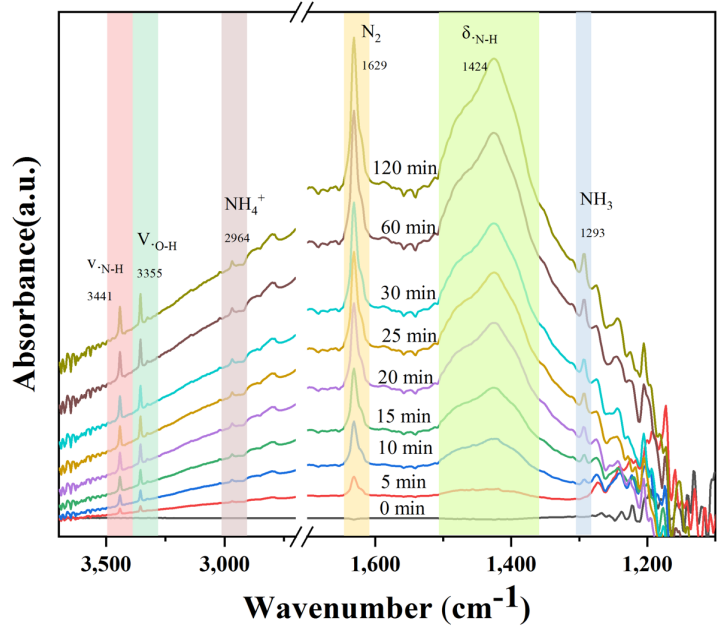

Figure 11. In-situ FTIR spectra of 0.25Fe-WO$_3$ photocatalytic nitrogen fixation

## 3.5 Interactions between light and photocatalyst

The interactions between light and the synthesized Fe-WO$_3$ photocatalytic particles is critical to study their photocatalytic performance. Two key features are of special interests: (i) the absorption cross section which represents how much energy from the sun light can be taken by the photocatalytic particles; and (ii) the surface charge distriubtion on the photocatalic particle boundary which might make siginifcant contribution to the electron–hole separation on the particle surface for the photocatalytic process. Such features are highly related to the material properties and the morphology of particles.

The numerical modelling was performed to investigate the absorption and surface charge distribution of a single 0.25Fe-WO$_3$ photocatalytic particle under the illumination of sunlight in water by using the field-only surface integral method[47-51]. As shown in Figure 12c, the typical morphology of our synthesized 0.25Fe-WO$_3$ photocatalytic particle is a hexagon disk with irregular bumps on the particle surface. In the simulations, the edge length of the hexagonal particle was set as 300 nm and the thickness of the particle was chosen as 100 nm. The irregular bumps on the particle surface was introduced by using the idea of Gaussian random particle[52]with the relative standard derivation of 80 nm and the corresponding angle of 5°. The refractive index of water was set as $n_{\text{water}} = 1.33$. The effective refractive index $n_{\text{eff}}$ and extinction coefficient $k_{\text{eff}}$ of the 0.25Fe-WO$_3$ photocatalytic particle under different wavelengths, as plotted in Figure 12a, were obtained by using the mole averaged value of the the refractive index and extinction coefficient of WO$_3$[53,54] and those of Fe$_2$O$_3$[55]which leads to $n_{\text{eff}} = 0.75 n_{\text{WO}_3} + 0.25 n_{\text{Fe}_2\text{O}_3}$ and $k_{\text{eff}} = 0.75 k_{\text{WO}_3} + 0.25 k_{\text{Fe}_2\text{O}_3}$, respectively.

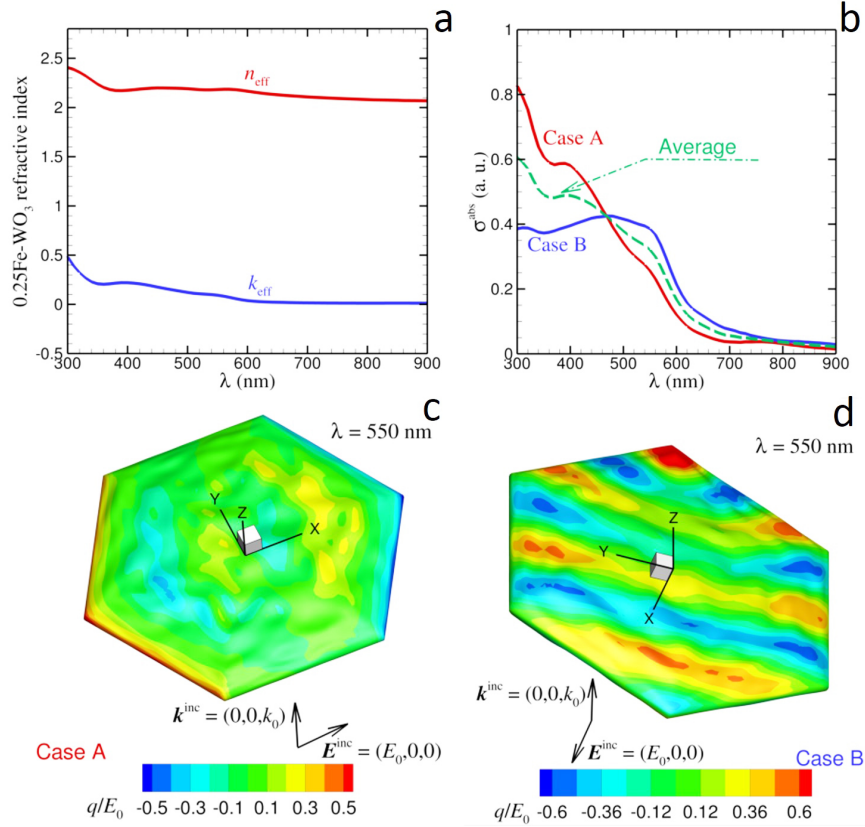

Figure 12. (a) Effective refractive indices and extinction coefficients of 0.25Fe-WO$_3$ photocatalytic particle under different wavelengths. (b) Absorption cross sections of a 0.25Fe-WO$_3$ disk particle under the illumination of sunlight when the disk is perpendicular to (Case A) and parallel to (Case B) light propagation. Distribution of surface charge density $q$ on a 3D 0.25Fe-WO$_3$ disk particle at wavelength $\lambda = 550$ nm when the disk is (c) perpendicular to and (d) parallel to the incoming light.

To numerically mimic the experiments, plane electromagnetic wave with time harmonic $\exp(-i\omega t)$ was used to represent the sunlight in which $\omega$ is the angular frequency of the light and i is the imaginary unit. The spectrum of wavelength, $\lambda = 2\pi c/\omega$ where $c$ is the speed of light, around the visible light was screened from 300 nm to 900 nm with a step of 10 nm. It was set that the incoming plane travels along $z$-direction as $\boldsymbol{k}^{\text{inc}} = (0,0,k_0)$ with $k_0 = 2\pi/\lambda$ being the wavenumber and its electric field polarizes along $x$-direction as $\boldsymbol{E}^{\text{inc}} = \boldsymbol{E}_0\exp(ik_0 z)$ with $\boldsymbol{E}_0 = (E_0, 0, 0)$, as illustrated in Figure 12c,d. The scattered electromagnetic fields by the 0.25Fe-WO$_3$ photocatalytic particle in water, $\boldsymbol{E}^{\text{sca}}$ and $\boldsymbol{H}^{\text{sca}}$, and the transmitted field within the particle, $\boldsymbol{E}^{\text{tra}}$ and $\boldsymbol{H}^{\text{tra}}$, were calculated by using the field-only surface integral method[47-50].

Two typical cases when the 0.25Fe-WO$_3$ disk particle is perpendicular to (Case A) and parallel to (Case B) the light propagation direction were under consideration in the modelling work. In Figure 12b, the variation of absorption cross section $\sigma^{\text{abs}}$ for Case A and Case B across the wavelength spectrum of sunlight is presented. The definition of $\sigma^{\text{abs}}$ is

$$\sigma^{\text{abs}} = \int_S \frac{1}{2}\text{Real}[\boldsymbol{E}^{\text{tot}} \times (\boldsymbol{H}^{\text{tot}})^*] \, \text{d}\boldsymbol{S}$$

in which $\boldsymbol{E}^{\text{tot}} = \boldsymbol{E}^{\text{inc}} + \boldsymbol{E}^{\text{sca}}$ and $\boldsymbol{H}^{\text{tot}} = \boldsymbol{H}^{\text{inc}} + \boldsymbol{H}^{\text{sca}}$, superscript * indicates the conjugate of the field values, and S is a surface that encloses the particle. In our calculations, we set *S* is a spherical surface with radius of 1.0 μm. The obvious absorption of sunlight by the 0.25Fe-WO$_3$ photocatalytic particle as shown in Figure 12b indicates that the newly synthesized material is an efficient semiconductor that is suitable to harvest solar energy to for electron–hole transfer as a potentially effective photocatalyst material. To better represent the pratical exmperimental system which contains many particles with different orientations, the average (arithmetic mean) of the absorption cross sections of these two typical cases was taken as shown by the dash line in Figure 12b. When compared to the absorption measured in the experiments given in the green line in Figure 5a, the trend from the modelling of the averaged aborption across the sunlight spectrum is in good agrenment with that obtained in experiments.

Figure 12c,d show the distribution of the surface charge density, *q*, on the internal side of the 0.25Fe-WO$_3$ photocatalytic particle surface for Case A and Case B where $q = \boldsymbol{E}^{\mathrm{tra}} \cdot \boldsymbol{n}$ with $\boldsymbol{n}$ being the unit normal vector on the surface pointing into the particle. The distribution of *q* on the particle surface can illustrate the potential movement of electrons and holes. As shown in Figure 12c, for Case A when the 0.25Fe-WO$_3$ disk particle is perpendicular to the wave propogation, high concentration of surface electric charges accumulate at the thin disk edges and corners with high curvatures which introduce strong gradient of electric surface charge around those areas. Also, for Case B in Figure 12d when the 0.25Fe-WO$_3$ disk particle is parallel to the incoming sunlight, it can be seen that, besides the oscillated distribution of *q* across the whole particle surface, the local geometric feature, such as high curvature due to irregular surface bumps, can induce obvious gradient of electric surface charges. These strong field gradients are believed to be benefical to the electron–hole separation for the photocatalytic performance.

## 4. Conclusion

A series of oxygen vacancy iron-doped WO$_3$ photocatalysts were designed and prepared by high-temperature calcination. This photocatalyst with a large number of vacancies can effectively adsorb and fix nitrogen from air directly. The incorporation of Fe can not only improve the light absorption efficiency, but also improve the ability of transporting electrons. When the amount of Fe doping is 0.25 times that of W, the nitrogen fixation efficiency is 4 times that of pure WO$_3$. The optium reaction conditions of 2 mg catalyst in 10 ml water at 55℃ were optimized by orthogonal experiments. The maximum nirogen fixation rate of 0.25Fe-WO$_3$ could reach 477 μg·g$_{\mathrm{cat}}^{-1}$·h$^{-1}$ from air under sunlight. This fixation process was confirmed by in situ infrared. Meanwhile, modelling on the interactions between light and the photocatalyst was carried out to study the distribution of surface charge and validate the light absorption of the photocatalyst. This work provides an easy, efficient and affordable iron-doped tungsten oxide photocatalyst for fixing nitrogen, which can pratically contribute to the carbon-neutrally sustainable development.

## Acknowledgement

We are thankful to the National Natural Science Foundation of China (21978061), Zhejiang Provincial Natural Science Foundation of China (LY19B060007) and Zhejiang Key Laboratory of Green Pesticides and Cleaner Production Technology for providing financial support. Q. Sun acknowledges the support by the Australian Research Council (ARC) through grants DE150100169, FT160100357 and CE140100003. This research was undertaken with the assistance of resources from the National Computational Infrastructure (NCI Australia), an NCRIS enabled capability supported by the Australian Government.# References

[1] S. Wang, F. Ichihara, H. Pang, H. Chen, J. Ye, Nitrogen Fixation Reaction Derived from Nanostructured Catalytic Materials, Adv. Funct. Mater. 28 (2018) 1803309.

[2] Z. Zhao, H. Ren, D. Yang, Y. Han, J. Shi, K. An, Y. Chen, Y. Shi, W. Wang, J. Tan, X. Xin, Y. Zhang, Z. Jiang, Boosting Nitrogen Activation via Bimetallic Organic Frameworks for Photocatalytic Ammonia Synthesis, ACS Catal. 11 (2021) 9986-9995.

[3] W. Kong, Z. Liu, J. Han, L. Xia, Y. Wang, Q. Liu, X. Shi, Y. Wu, Y. Xu, X. Sun, Ambient electrochemical $N_2$-to-$NH_3$ fixation enabled by $Nb_2O_5$ nanowire array, Inorg. Chem. Front. 6 (2019) 423-427.

[4] A. Bagger, H. Wan, I.E.L. Stephens, J. Rossmeisl, Role of Catalyst in Controlling $N_2$ Reduction Selectivity: A Unified View of Nitrogenase and Solid Electrodes, ACS Catal. 11 (2021) 6596-6601.

[5] F. Schendzielorz, M. Finger, J. Abbenseth, C. Wurtele, V. Krewald, S. Schneider, Metal-Ligand Cooperative Synthesis of Benzonitrile by Electrochemical Reduction and Photolytic Splitting of Dinitrogen, Angew. Chem. Int. Ed. 58 (2019) 830-834.

[6] Y.-H. Liu, C.A. Fernández, S.A. Varanasi, N.N. Bui, L. Song, M.C. Hatzell, Prospects for Aerobic Photocatalytic Nitrogen Fixation, ACS Energy Lett. 7 (2021) 24-29.

[7] H. Sun, F. Quan, C. Mao, L. Zhang, New Strategies for Nitrogen Fixation and Pollution Control, Chinese J. Catal. 39 (2021) 3199-3210.

[8] A.W. Bataller, R.A. Younts, A. Rustagi, Y. Yu, H. Ardekani, A. Kemper, L. Cao, K. Gundogdu, Dense Electron-Hole Plasma Formation and Ultralong Charge Lifetime in Monolayer $MoS_2$ via Material Tuning, Nano Lett. 19 (2019) 1104-1111.

[9] W. Zhang, Y. Fu, Q. Peng, Q. Yao, X. Wang, A. Yu, Z. Chen, Supramolecular preorganization effect to access single cobalt sites for enhanced photocatalytic hydrogen evolution and nitrogen fixation, Chem. Eng. J. 394 (2020) 12482.

[10] P. Li, Z. Zhou, Q. Wang, M. Guo, S. Chen, J. Low, R. Long, W. Liu, P. Ding, Y. Wu, Y. Xiong, Visible-Light-Driven Nitrogen Fixation Catalyzed by $Bi_5O_7Br$ Nanostructures: Enhanced Performance by Oxygen Vacancies, J. Am. Chem. Soc. 142 (2020) 12430-12439.

[11] B. Sun, Z. Liang, Y. Qian, X. Xu, Y. Han, J. Tian, Sulfur Vacancy-Rich O-Doped 1T-$MoS_2$ Nanosheets for Exceptional Photocatalytic Nitrogen Fixation over CdS, ACS Appl. Mater. Interfaces 12 (2020) 7257-7269.

[12] S. Zhang, L. Zhang, S. Fang, J. Zhou, J. Fan, K. Lv, Plasmonic semiconductor photocatalyst: Non-stoichiometric tungsten oxide, Environ. Res. 199 (2021) 111259.

[13] Gerhard N. Schrauzer, Norman Strampach, Liu Nan Hui, Miles R. Palmer, A. J. Salehi, Nitrogen photoreduction on desert sands under sterile conditions, Chemistry 80 (1983) 3873-3876.

[14] G. N. Schrauzer, a.T.D. Guth, Photolysis of Water and Photoreduction of Nitrogen on Titanium Dioxide, J. Am. Chem. Soc. 99 (1977) 7189-7193.

[15] A. Krumina, D.A. Rasmane, D. Jankovica, J. Grabis, R. Drunka, Synthesis, photocatalytic properties and morphology of various $TiO_2$ nanostructures modified with gold, P. Est. Acad.